\documentclass[aps,pre,twocolumn,floatfix,showpacs,showkeys]{revtex4}
\usepackage{amsmath,amsfonts,amssymb,mathrsfs,graphicx}
\begin{document}
\title{Neural Networks with Quantum Gated Nodes}

\author{Fariel Shafee}
\affiliation{ Department of Physics\\ Princeton University\\
Princeton, NJ 08540\\ USA.} \email{fshafee@princeton.edu }

\begin {abstract}
We study a quantum neural network with superposed qubits replacing
classical neurons with deterministic states, and also with quantum
gate operators in place of the classical action potentials observed
in biological contexts. With our choice of logic gates
interconnecting the neural lattice, we find that the state of the
system behaves in ways reflecting both the strength of coupling
between neurons as well as the initial conditions, and depending on
whether there is a threshold for emission from excited to ground
state, the system shows either chaotic oscillations or coherent ones
with periodicity that depends on the strength of coupling in a
unique way. The spatial pattern of the initial input affects the
subsequent dynamic behavior of the system in an interesting
unambiguous way, which indicates that it can serve as a dynamic
memory system analogous to biological ones, but with an unlimited
lifetime.

\end{abstract}

 \vspace*{1cm}

\maketitle

\section{INTRODUCTION}

Quantum computers have recently become a topic of elaborate
research. It has been shown \cite{NC1} that a quantum computer can
compute some NP hard problem much faster than a classical computer
because the memory elements can simultaneously hold multiple
information and the processor can operate in parallel on many
qubits.  Quantum entanglement between the elements permits the use
of different kinds of algorithms \cite{SH1, GR1} for such quantum
computers as well, which too are now being explored extensively.

Closely related to computing is the process of memory and pattern
recognition. Biologically, the human brain has been modeled as a
network of neurons firing signals with different time sequence
patterns corresponding to different input signals. Hebbian learning
is achieved by taking into account the plasticity of the weights
with which the neurons are connected to each other. A dynamic type
of memory system in a classical network with integrate-and-fire
neurons was first presented by Hopfield and Herz \cite{JH1} which
was later extended to the more complicated case of a still classical
model with nonzero widths of the action potentials \cite{FS1} with
some features similar to the zero-width case and other novel
features related to the finite width.

An effort to enter the quantum realm with such dynamic networks was
made in the semiclassical version reported in \cite{FS2}. In all
these works the periodicity and the pattern of collective and
firings of the dynamical neural networks acted on by the environment
and also with nearest neighbor action between the neighboring
neurons were studied in detail. However, in the previous
semiclassical neural network model the analogies retained with the
classical case took it somewhat away from the current focus in
quantum computing, where quantum gates replace arbitrary
time-dependent potentials so that there is a well-defined unitary
operator representing the transitions at the nodes of the network.

In the present work, we investigate a quantum machine with such
easily referrable and repeatable quantum gates, which at the same
time also most closely resembles our previously studied models, so
that we can understand aspects of the differences between a
classical or a semiclassical model having given potentials between
the nodes, with a quantum computer-like gated system, where the
building block is the quantum transition operation representing a
logical operation, not a general potential.

Whereas a quantum computer would be useful for calculating
quantities, with an input operated on by the processor and producing
a quantitative output, a quantum neural network may possibly also be
used for the purpose of enriched learning of a variety different
from the Hebbian learning of classical networks. This possibility
has not yet been fully explored, though Altaisky \cite{AL1} has made
some preliminary investigations into a single quantum perceptron.
Learning is of course an irreversible process, and the unitarity of
the operators involved in a net does not permit an irreversible
change of the system in the ordinary sense. There have been many
attempts \cite{OT1} to explain such a change using decoherence at
the output with reversibility of the intermediate processes in a
quantum computer. This approach has mostly been suggested for
quantum computers. For quantum neural networks usually the
transition to a certain eigenstate is inserted in an {\it ad hoc}
manner, as in the work of Altaisky and also that of Zak et al
\cite{ZK1}.

In this work we shall not go into the details of the complications
of the separation of the quantum processing followed by a classical
output. We shall instead try to mimic as nearly as possible our
previous biologically motivated model of an integrate-and-fire
neural network, because biological or quasi-biological models are
always a fascinating benchmark for the comparison of any system that
aims at intelligence-like functions. However, in this case we have,
in place of deterministic neurons, qubits being interconnected to
nearest neighbors by quantum gates with well-defined operatorial
roles .

\section{THE QUANTUM NEURAL NETWORK MODEL}

In the integrate-and-fire model a neuron receives a current from the
fired neighbors, and when its own potential exceeds the threshold it
too fires, feeding its own neighbors.  We have studied the effect of
the finite duration of the signal from a neuron to the neighbor. In
a quantum process all transitions of the neurons must be designated
by unitary operators. So in place of the firing of a neuron we have
a less spectacular unitary transformation that simply performs a
sort of rotation of the state vector or the qubit.

In principle this operation should involve time too, and we should
write:
\begin{equation} \label{eq1}
|t \rangle = U(t,t_0)|0 \rangle
\end{equation}

to indicate the transformation of a neuron from time $t_o$ time $t$.
For small time changes it is usually possible to write:

\begin{equation} \label{eq2}
U(t+dt, t)  = U(t, t)  +  i dt  H
\end{equation}

So that
\begin{equation} \label{eq3}
d |t\rangle   = i dt H  |t\rangle
\end{equation}

in the lowest order, with a hermitian operator $H$, usually the
Hamiltonian.

In quantum computing it has been shown that a complete set of
unitary operators exist to express the classical logical operations
such as NOT, AND or XOR. These may make use of Hadamard gates,
phase-change gates or controlled-NOT (c-NOT) gates. These gates may
be combined to give entanglement between different nodes, e.g. the
c-NOT or the Toffoli gate, which is a kind of adder.

It is not necessary for the whole network to be completely
entangled by the basic operators of the net to form a useful
network. It is known that we can have pair-wise entanglements at
the lowest nontrivial level. However, even if we entangle only
nearest neighbors, the entanglement may spread throughout the net
after successive operations. The process is similar to obtaining a
dense matrix from the multiplication of a large number of sparse
matrices with nonzero elements at different positions.

We postulate the following physical model:

\par
$1.$  the neurons represent  qubits;
\par
$2.$  an excited neuron $|1\rangle$  will turn on a neighbor in a
ground state $|0 \rangle$;
\par
$3.$ an excited state will make an excited neighbor 'fire' and go
down to the ground state (induced emission) ;
\par
$4.$  the excited state itself will go down to the ground state
after exciting the neighbors;
\par
$5.$ an unexcited neuron does nothing to itself or any neighbor.

Postulates 2, 3 and 5 can be satisfied by a c-NOT gate, with the
first neuron serving as the controller, and operating on its
neighbor. With a square lattice we consider for simplicity there
are four neighbors for each neuron, so that in place of c-NOT
gates we shall need $c-NOT^4$ gates, i.e. one controller flipping
all four neighbors if it is in state $|1 \rangle$ and doing
nothing if it is in state $|0 \rangle$.

Postulate 4 can be satisfied by using an AND gate connection every
neuron with a common $|0 \rangle$ sate after the $c-NOT^4$ gate.

For any particular neighbor the c-NOT gate can be represented by
\begin{equation} \label{eq4}
 U =  \left(   \begin{array} {cc}
 1 & 0 \\
0 & \sigma_1
\end{array} \right)
\end{equation}

where $\sigma_1$ is the flipping Pauli matrix. Eq.\label{eq4}
represents a hermitian operator.  We, however, connect the c-NOT
matrices using a weight factor, $\epsilon$ to represent the strength
with which a neuron can affect its neighbor. This weight factor is
not the same as the weight factors in a classical neural network,
which must sum up to one, but just a measure of the strength with
which the neurons are able to affect the neighbors.

The $.AND. |0\rangle$ can be represented by another hermitian
operator in the qubit space:

\begin{equation} \label{eq5}
U ^{\prime} = \left( \begin {array} {cc}
0 & 0 \\
0 & 1   \end{array} \right)
\end {equation}

where the sequence of states in the rows and columns are, as usual
$|1\rangle|1\rangle$, $|1\rangle|0\rangle$, $|0\rangle|1\rangle$
and $|0\rangle|0\rangle$, the first being the controlling state.

Although hermitian, the AND operator is obviously not unitary.  The
nonunitarity of the AND operator is responsible for the collapse of
the state to the ground state after it has reached a threshold. This
situation is not the standard quantum computer situation, where all
gates are required to be strictly unitary, but is a mix between
unitary connections among neurons and a nonunitary collapse.  The
introduction of unitary rotations of the neurons in the qubit space
makes our device fast and efficient, and the collapse to the ground
state by the AND operator lets us mimic the classical neural network
model as closely as possible.  However, The collapse of the state to
the ground state simply restarts the rotational activity of a neuron
when connected to an excited neighbor.  So this situation is not
quite the same as decoherence when in contact with nature, and
states are collapsed probabilistically to one of the superposed
states and delete all memory.  Rather, this collapse is a
conditional collapse to a specified state, and the timing of the
collapse holds information about when that specific threshold was
reached to enable a time sequence pattern.

So we get at each node the controlling qubit remaining unchanged due
to its own action:
\begin{equation}  \label{eq6}
\left(  \begin{array} {c} c \\ s \end{array} \right)\rightarrow
\left(
\begin{array} {c} c \\ s \end{array} \right)
\end{equation}

and the following change for the neighbors receiving signal from
it, i.e. operated on by the $c-NOT^4$ gates:

\begin{equation}  \label{eq7}
\left(  \begin{array} {c} c^{\prime} \\ s^{\prime} \end{array}
\right)\rightarrow \left(  \begin{array} {c} c^{\prime} \\
s^{\prime}
\end{array} \right)
+\epsilon \left(
\begin{array} {c} -s^{\prime}.c \\ c^{\prime}.c \end{array}
\right)
\end{equation}

Since the small $\epsilon$ approximation of the unitary operators
are not themselves unitary, in simulation it is necessary to
renormalize each qubit at each step.

\section{RESULTS OF SIMULATION}
As in the classical cases in previous works, for ease of comparison
we constructed a 40X40 network of qubits with periodic boundary
conditions, so that it behaves in some ways as a much larger
lattice. In the quantum case of course in reality a large lattice
would very difficult to realize in the laboratory at the present
stage of technology, but for such a theoretical study it makes no
difference.

We put the input data on qubits in the periphery and made the
inside neurons either all random or all zero $[(0, 1)] $. Then we
updated the qubits according to Eq.\ref{eq6} and \ref{eq7}.

A large number $(40,000)$ of time steps were chosen and various
$\epsilon$  values. This parameter may be interpreted as the
strength of coupling of the neurons, but as it occurs together
with $dt$, it may also indicate the width of the pulse at each
time step.

In our first model we did not use a threshold for the firing of
the neurons, so that each neuron was allowed to go up to its full
qubit value of $(1, 0)$ with $c= +1$ or $-1$. It is interesting to
observe that in this case  there is no well-defined periodicity,
either for any single neuron or for the average neuron (i.e. the
sum) in the network (Figs. 1, 2). Though all neurons do indeed go
through the $(1, 0)$ to $(0, 1)$ cycles, the oscillations are
aperiodic. This may be because the exact equation of motion
coupling the nearest neighbor neurons becomes insoluble in terms
of periodic functions.
\begin{figure}[ht!]
\includegraphics[width=8cm]{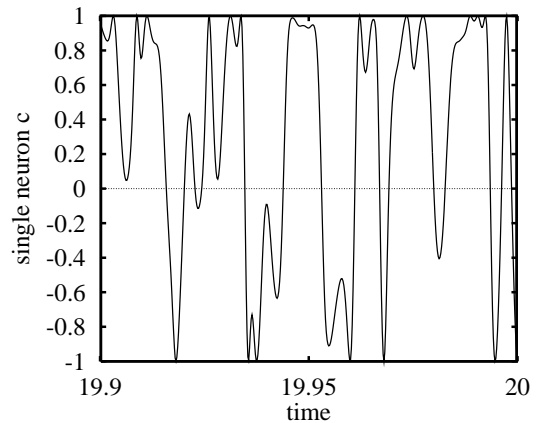}
\caption{\label{fig1} Oscillations of the c part of a qubit for a
non-cutoff model with $\epsilon= .01$. There is no fixed
periodicity.}
\end{figure}
\begin{figure}[ht!]
\includegraphics[width=8cm]{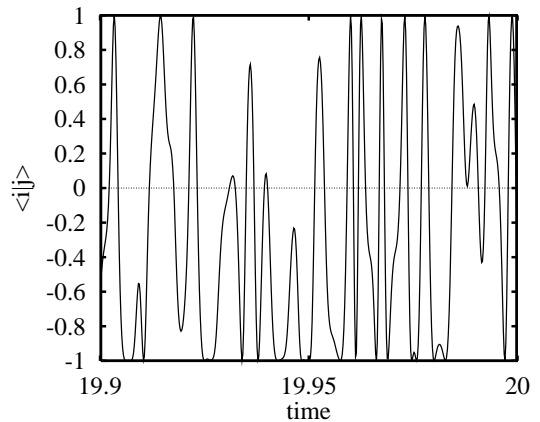}
\caption{\label{fig2} Correlation between two qubits for no
threshold case with $\epsilon= 0.01$, $\langle10,10| 20,21\rangle$,
where the qubits are located by their $(x, y)$ coordinates.}
\end{figure}

We also experimented with a slightly different version of the model
of the network more akin to the classical one. Here we introduced a
threshold for the excited part of the neuron, which when crossed,
causes the qubit to jump to the ground state $(0, 1)$, i.e. if  $c >
c_thres$, any $(c, s)$ makes a transition to the $(0, 1)$ state.
This may be considered to be due to emission of energy by an excited
element on reaching a threshold. So, in this case the
$.AND.|0\rangle $ operation works on reaching the threshold.  A
notable difference between the two models is that whereas in the
first model, the AND gate brought back the qubit slowly to the
ground state, in the second model, the qubit was collapsed to the
ground state immediately.

We found more interesting results with this model. In this case we
get periodic oscillations of the system, with all neurons almost
in the same phase. We put the threshold at $0.7$ which is just
below $1/\surd 2$, because this seems to be critical threshold
that gives regular oscillations.

We believe this happens because here the cut-off effectively
serves to truncate  the complicated coupled behavior of the
system, reducing it to a simpler periodic system,  just as the
truncation of a transcendental function by a polynomial with a
finite number of terms provides it with a simpler behavior. For
example, the function:

\begin{equation}  \label{eq8}
F(t) = \cos (\theta + \epsilon \sin \theta)
\end{equation}
with $\theta = \omega t $ assumes the periodic form
\begin{equation} \label{eq9}
F(t) = \cos [ (1+ \epsilon) \omega t]
\end{equation}

for small enough $\epsilon$  only, but has a more complicated
behavior at large $\epsilon$. Obviously this requires more thorough
and careful study.

However, the other reason for this behavior might be that the rate
at which the AND operator acted was different from the rate at which
the c-NOT operator acted on the system, and the coupling of the two
incongruous frequencies yielded a complex pattern.

We note that these oscillations are seen in the behavior of a
single neuron (Fig. 3), or the sum of all neurons of the system,
or even in the correlation $\langle i|j \rangle$ between neuron
$|i\rangle$ and neuron $|j\rangle$ (Fig. 4).

\begin{figure}[ht!]
\includegraphics[width=8cm]{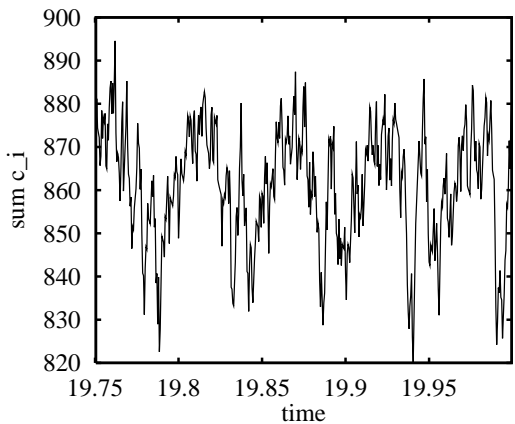}
\caption{\label{fig3}Oscillation of the c parts summed of all qubits
in the network for $threshold = 0.7$, $\epsilon= 0.01$. All
boundaries excited initially.}
\end{figure}

\begin{figure}[ht!]
\includegraphics[width=8cm]{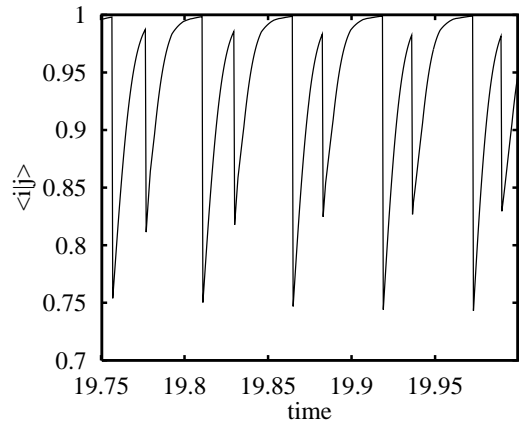}
\caption{\label{fig4}Correlation $\langle 10,10|20,21\rangle$ for
the above case.}
\end{figure}

Most interestingly, it appears that for large $\epsilon$ ($ >
0.7$), if we put the initial signal only at two parallel sides of
the square, there is no oscillation (Fig. 5), but a static
asymptotic state, which is quickly reached, but if we put the
signal on all four sides, then periodic oscillations continue with
changed frequency.

\begin{figure}[ht!]
\includegraphics[width=8cm]{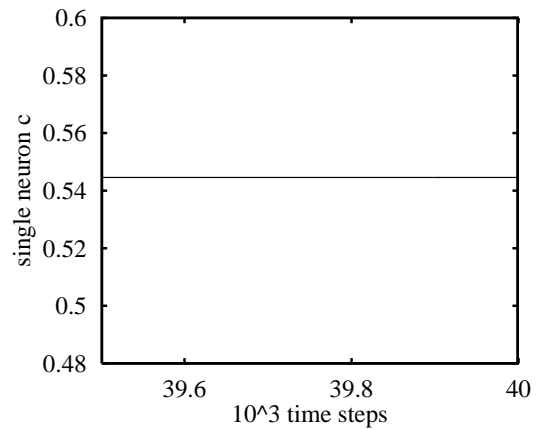}
\caption{\label{fig5}$c$ of a single neuron for $\epsilon = 0.8$,
only $x$ sides excited initially. There are no oscillations
asymptotically (we have plotted only after 39500 time steps).}
\end{figure}

It is possible that the lack of signal in the orthogonal direction
allows the neurons in the net the extra freedom to adjust
themselves to fixed static states in the direction of the signal.
This is reminiscent of the one-dimensional Ising model having a
trivial phase transition. When signals arrive from both $x$ and
$y$-directions, presumably the attractor for the system becomes
dynamic, as it tries to adjust in both directions, but cannot find
a static equilibrium state.

\section{ CONCLUSIONS}

We have shown that a quantum neural network  similar to the
integrate-and-fire neuron network in some ways can be constructed
with quantum elements, consisting of qubit nodes connected by c-NOT
and AND gates. However, the quantum gated system also has
significant differences.

We have found the interesting property that, with no threshold the
system converges to a dynamic state with no fixed period and no
phase locking, similar to a chaotic system, but with an average
behavior which is not entirely chaotic.

\begin{figure}[ht!]
\includegraphics[width=8cm]{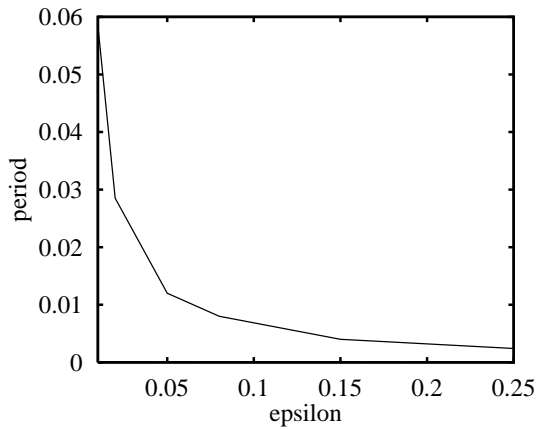}
\caption{\label{fig6}Variation of periodicity with $\epsilon$ for
excitations from all sides.}
\end{figure}

With an assigned threshold that takes an excited qubit to the zero
state, we see dynamic oscillations of the system. The period is
almost inversely proportional to the coupling strength (Fig. 6), but
shows nonlinearity for strong coupling. For quite strong coupling,
if there are initial excitations only in one direction, the system
seems to converge rapidly to a static attractor, but with
excitations from both directions of the square lattice, dynamic
oscillations continue. This is a very significant difference in the
mapping of the initial spatial pattern into dynamic patterns of the
system that promises to make such a system a useful device for
conversion of static patterns into dynamic memory.

In the case of fixed period oscillations, the correlation between
neurons as measured by the overlap of the two qubits, $\langle i|j
\rangle$ also shows periodic time dependence. Interestingly ,we
have found that despite the simplicity of the model of this
quantum neural network with c-NOT gates, it can hold dynamic
memories of the input indefinitely.

Obviously, with complex phases in the coupling the pattern
generated may be more interesting, also if there is a dynamic
external agent affecting the peripheral neurons, and not just an
initial input. We can also introduce delay lines between the
neurons to introduce a time scale.

\end{document}